\pdfoutput=1

\documentclass[11pt]{article}

\usepackage[]{EMNLP2022}

\usepackage{times}
\usepackage{latexsym}

\usepackage[T1]{fontenc}

\usepackage[utf8]{inputenc}

\usepackage{microtype}

\usepackage{inconsolata}

\usepackage{algorithm}
\usepackage{algorithmic}

\usepackage[textsize=scriptsize]{todonotes}

\usepackage{cleveref}

\crefname{section}{§}{§§}

\usepackage{multicol}
\usepackage{multirow}
\usepackage{newfloat}
\usepackage{listings}
\usepackage{booktabs}
\usepackage{tikz}
\usepackage{calc}
\usepackage{amssymb}
\usepackage{bbm}

\title{Execution-based Evaluation for Data Science Code Generation Models}

\author{
Junjie Huang$^{1}\thanks{\ \ Work done during internship at Microsoft Research Asia.}$\hspace{0.2em},
 Chenglong Wang$^{3}$,
 Jipeng Zhang$^{5*}$, 
 Cong Yan$^{3}$,
 Haotian Cui$^{6*}$,\\
 {\bf Jeevana Priya Inala$^{3}$,
 Colin Clement$^{4}$,
 Nan Duan$^{2}$,
 Jianfeng Gao$^{3}$}\hspace{0.5em}\\
$^1$ Beihang University  \quad $^2$ Microsoft Research Asia   \quad $^{3}$Microsoft Research Redmond \\ 
$^4$ Microsoft \quad
$^5$ Hong Kong University of Science and Technology \quad 
$^6$ Toronto University \\
{\tt $^{1}$huangjunjie@buaa.edu.cn, } \\ 
{\tt $^{2,3,4}$\{chenwang,coyan,janala,coclemen,nanduan,jfgao\}@microsoft.com  } \\
{\tt  $^{5}$jzhanggr@connect.ust.hk, $^{6}$ht.cui@mail.utoronto.ca}
}

\begin{document}
\maketitle
\begin{abstract}

Code generation models can benefit data scientists' productivity by automatically generating code from context and text descriptions.
An important measure of the modeling progress is whether a model can generate code that can correctly execute to solve the task.
However, due to the lack of an evaluation dataset that directly supports execution-based model evaluation, existing work relies on code surface form similarity metrics (e.g., BLEU, CodeBLEU) for model selection, which can be inaccurate. 

To remedy this, we introduce ExeDS, an evaluation dataset for execution evaluation for data science code generation tasks. ExeDS contains a set of 534 problems from Jupyter Notebooks, each consisting of code context, task description, reference program, and the desired execution output.
With ExeDS, we evaluate the execution performance of five state-of-the-art code generation models that have achieved high surface-form evaluation scores. 
Our experiments show that models with high surface-form scores do not necessarily perform well on execution metrics, and execution-based metrics can better capture model code generation errors.
\footnote{Source code and data can be found at \url{https://github.com/Jun-jie-Huang/ExeDS}.}

\end{abstract}

\section{Introduction}\label{sec:intro}
\begin{figure}[t]
     \centering
     \includegraphics[width=0.475\textwidth]{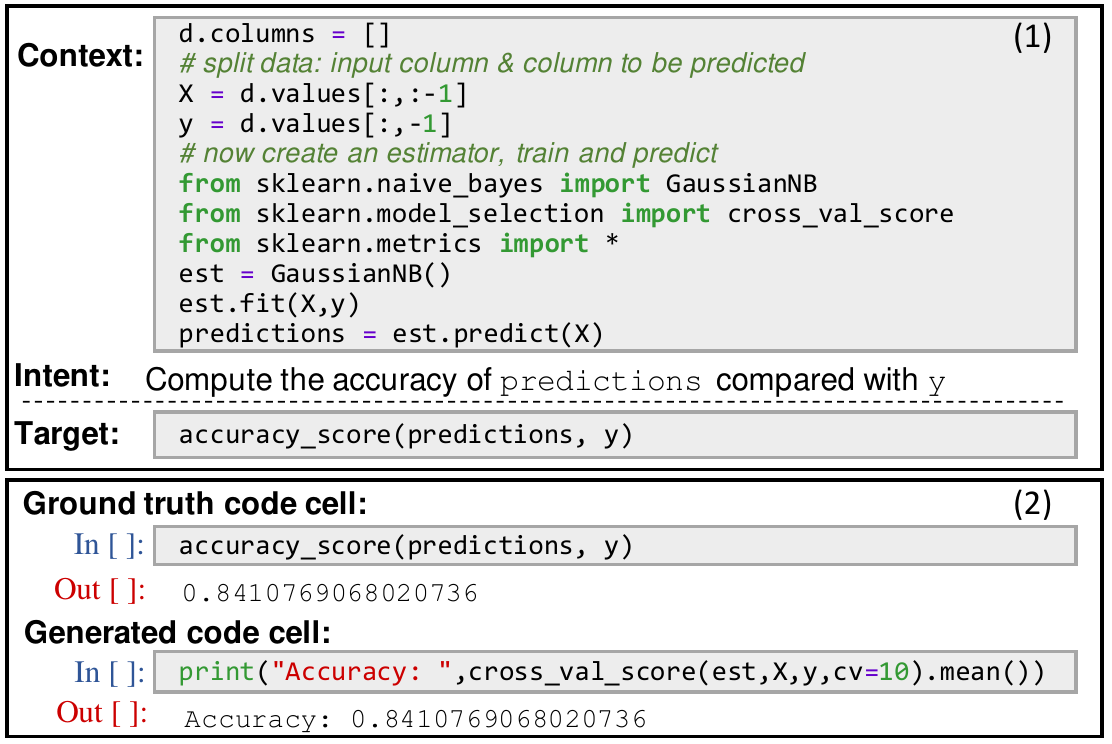}
     \caption{An example from ExeDS. The first block describes the task of data science code generation with code context and NL intents.
     The second block compares the code and output of reference and generation.
     }
     \label{fig:example}
\end{figure}

Code generation models \cite{ Chen2021CodeX, tunstall2022codeparrot} have shown promising results to improve developer productivity by generating code from natural specifications \cite{Le2020ACMSurvey, AlHossami2022Asurvey}.
These promising results also bring interest to code generation for data scientists, who program data analysis scripts in interactive notebook environments like Jupyter Notebooks \cite{Kluyver2016JupyterNotebook} where programs are written interactively in loosely organized program cells 
(Figure \ref{fig:example} (1)).
This domain and style differences motivates new modeling resources, e.g., new datasets (e.g. JuiCe \cite{Agashe2019JuICe} 
) and models (e.g. JuPyT5 \cite{Chandel2022JuPyT5}) specific to data science tasks.

However, we still lack a good methodology to evaluate data science (DS) code generation models. 
JuiCe dataset uses the BLEU \cite{Papineni2002BLEU} and Exact Match (EM), the prevailing metrics in code generation, to measure semantic similarity between the generated and reference code. 
However, these two surface-form metrics have limitations:
the former neglects code syntactic features and the latter is too strict \cite{Ren2020CodeBLEU}. 
Execution-based metrics are another widely accepted line of metrics in general software engineering (SE) domain, where the correctness of generated functions is determined by whether the outputs are consistent with oracle input-output data/unit tests. 
For DS problems, however, collecting an executable dataset and performing execution-based evaluation are challenging. 
DS notebooks usually do not come with their own set of unit tests and existing datasets like JuiCe do not track the input data (such as tables) needed to run the notebooks. 
In addition, the outputs from notebook cells are often not "pure" values (e.g., numbers, strings, or lists) like the outputs of the functions in SE problems. The DS notebook cell outputs are meant for human understanding and hence, may contain complex data structures (e.g., data frames, plots) accompanied with texts; 
thus simply checking whether outputs are the same is too strict to capture cases when the generated cell output is semantically correct but formatted differently from the reference 
(Figure \ref{fig:example} (2)).

In this paper, we provide a dataset for evaluating DS code generation models, dubbed ExeDS, which contains 534 data science problems built on JuiCe \cite{Agashe2019JuICe}.
We collect ExeDS by first crawling data dependencies from original GitHub repositories for the notebooks and filtering out notebooks with runtime errors; 
then, we curated 534 high-quality problems with sufficient code context and human-written natural language (NL) to describe tasks as the testset. 
With ExeDS, we can evaluate execution correctness by comparing outputs from generated code with desirable outputs.

We experiment with 5 existing code generation models on ExeDS to identify their execution performance.
Experiment results show that (1) models with high/low surface-form scores do not necessarily generate execution-correct code -- for example, while Codex \cite{Chen2021CodeX} is low in BLEU, it achieves high execution scores. 
(2) Execution-based metrics can better capture code errors which might be helpful for model improvements.

\section{Related Works}
Data science (DS) refers to the practice of analyzing data and acquiring insights with computational methods \cite{Donoho201750YO}. 
With the goal to improve productivity, there are increasing interests in building systems to solve a variety of DS tasks, including code synthesis \cite{Agashe2019JuICe}, code synthesis for visualization \cite{Chen2021plotcoder} and data preparation \cite{Yan2020AutoSuggest}, documentation \cite{Liu2021HAConvGNN, Wang2021GraphAugmentedCS}, etc.
In our work, we focus on code generation in DS, which generates code with code, NL and data context.

Code generation benchmarks are predominantly evaluated by matching code surface forms\cite{Papineni2002BLEU, Lin2004ROUGE, Ren2020CodeBLEU}. 
These datasets evaluate explicit code generation with different input specifications, including natural language \cite{Wang2015BuildingAS, Oda2015LearningTG, Zhong2017WikiSQL, Yin2018CoNaLa, Yu2018Spider, Lin2018NL2Bash}, unfinished code \cite{Iyer2018ConCode, lu2021codexglue}, and 
input-output examples \cite{Polosukhin2018NeuralPS, Zavershynskyi2018NAPS}.
However, surface form metrics are unable to assess code as  programmers, who focus on the functionality and execution correctness in practice.

Consequently, recent works turn to execution-based metrics instead, where the code would be correct if it passes a set of unit tests defined by humans \cite{NEURIPS2020UnitTranslation, Kulal2019SPoC, Austin2021MBPP, Chen2021CodeX, Hendrycks2021APPS}. 
However, the complex output data and scarcity of units tests in DS limit its application in DS code generation.
\citet{Chandel2022JuPyT5} explore applying unit tests in DS, but they only focus on educational problems.
Table \ref{tab:dataset-comparison} compares ExeDS with various related datasets.

\begin{table}[t]
  \centering
    \resizebox{0.475\textwidth}{!}{
    \begin{tabular}{lccl}
    \toprule 
        Dataset  & \multicolumn{1}{c}{\#} & Domain & \multicolumn{1}{c}{Evaluation} \\
    \midrule
    APPS \cite{Hendrycks2021APPS} & \multicolumn{1}{r}{10,000} & SE   & Unit Test \\
    MBPP \cite{Austin2021MBPP} & \multicolumn{1}{r}{974}  & SE   & Unit Test \\
    HumanEval \cite{Chen2021CodeX} & \multicolumn{1}{r}{164}  & SE    & Unit Test \\
    DSP \cite{Chandel2022JuPyT5} & \multicolumn{1}{r}{1,139}  & DS  & Unit Test \\
    JuiCe \cite{Agashe2019JuICe} & \multicolumn{1}{r}{1,981}  & DS   & Surface Form \\
    PlotCoder \cite{Chen2021plotcoder} &  \multicolumn{1}{r}{894} & DS    & Surface Form \\
    \midrule
    \textbf{ExeDS} (ours) & \multicolumn{1}{r}{534}  & DS  & Output Match \\
    \bottomrule
    \end{tabular}%
    }
  \caption{Comparisons of code generation testsets. 
  }
  \label{tab:dataset-comparison}
\end{table}%

\section{ExeDS for Execution Evaluation}
As mentioned in Section \ref{sec:intro}, the lack of executable environments for notebooks is a key limiting factor of execution-based model evaluation for data science tasks. 
Thus we first construct an evaluation dataset ExeDS and analyze its characteristics. Then describe the methods for execution evaluation.

\paragraph{Dataset Collection} ExeDS contains 534 problems with code context, NL task description, reference code and target execution output, which is built upon JuiCe \cite{Agashe2019JuICe} with 659K publicly available Python Jupyter notebooks from {GitHub}. 
We create ExeDS in the following steps.

\textbf{Step1: Crawling Data Context and Execution.} 
Programming problems in DS often deal with data, which are often stored in files (e.g., \texttt{.csv}) and loaded by code. 
Executing notebooks needs such data dependencies, which are not provided in JuiCe. 
Thus, we first crawl dependent data for notebooks from their GitHub repositories. 
Notebooks with inaccessible data or using libraries not present in Python standard library and default DS environment are removed.
With data dependency, we execute notebooks with a time limit of 1000 seconds per cell.
After execution, code cells have three types of outputs: (1) \textit{displaying data} with a figure; (2) \textit{execute result} with a textual execution output; and (3) \textit{stream output} with a printed textual output  through streams.
Since it's hard to compute figure similarities, 
in this paper, we only evaluate execution correctness on textual outputs and construct ExeDS with \textit{execute result} and \textit{stream output}.

\begin{table}[tbp]
  \centering
    \resizebox{0.40\textwidth}{!}{
    \begin{tabular}{lrl}
    \toprule
    Function Type & \multicolumn{1}{l}{\%} & Examples \\
    \midrule
    Data statistic & 40  & Avg., var.,  p-value, ... \\
    Explore data value & 19  & Min/max value, ...  \\
    Explore data property & 10  & Dtype, shape, ... \\
    Machine learning & 16  & Loss, train, predict, ... \\
    Simple math & 6   & Arithmetic, ...  \\
    Data changing & 5   & Sort, sample, ... \\
    Data displaying & 4   & Head/tail columns, ... \\
    \bottomrule
    \end{tabular}%
    }
  \caption{Function types of target code in ExeDS. }
  \label{tab:function-type}%
  \vspace{-3.5mm}
\end{table}%

\textbf{Step 2: Dataset Filtering and Intent Curation.} 
As some cells are overly complex for code generation, for simplicity, we remove examples with more than 5 lines or using customized methods in target code cells.  
To keep diversity, we downsample cells with frequent outputs, e.g. \texttt{df.summary()}, \texttt{df.info()},  \texttt{df.shape}, etc.
To ensure sufficient context is provided, we remove the target code whose variables are absent in the previous 5 cells.

Since some cells lack sufficient descriptions for the problems, for clarity, we recruit two university students with Python and notebook experience to manually write NL descriptions for each example.  
After viewing the context, target code and output, they are asked to write descriptions containing information in two aspects: (1) the functions of target code; (2) the instructions to print outputs. 
We discard examples that annotators feel hard to describe.

Finally, we obtain 534 problems from 278 notebooks for ExeDS, each with code context, NL description, target code, and desired execution output.

\paragraph{Dataset Statistics}
Table \ref{tab:function-type} shows the function types in ExeDS.  
We found the majority of target codes are computing statistics (40\%), exploring data value (19\%) or property (10\%), and for machine learning (16\%),
which are popular DS tasks.

Table \ref{tab:output-type} presents the types of execution output in all 534 problems. 
We find the majority of execution output are numbers, which is not surprising considering the fraction of data statistics and exploring data value in code functions. 
Also comparing numbers is less complicated than comparing other types of data like strings or data frames, which helps easier evaluation of execution outputs.

\begin{table}[thbp]
  \centering
    \resizebox{0.22\textwidth}{!}{
    \begin{tabular}{lr}
    \toprule
    Library & \multicolumn{1}{l}{\# problems} \\
    \midrule
    pandas & 534 \\
    numpy & 473 \\
    matplotlib & 431 \\
    sklearn & 287 \\
    seaborn & 211 \\
    scipy & 135 \\
    statsmodels & 57 \\
    math  & 46 \\
    datetime & 42 \\
    re    & 39 \\
    \bottomrule
    \end{tabular}%
    }
  \caption{Frequency of most common 10 libraries used  in 534 examples of ExeDS. }
  \label{tab:library}%
\end{table}%

Table \ref{tab:library} displays the most common libraries used in ExeDS. We find the majority of them use data science libraries and all of them use \texttt{pandas}, which indicates our focus on data science code generation.

\paragraph{Evaluation Metrics} In ExeDS, we measure the execution correctness by comparing the reference outputs with outputs from generated code, which is called output exact match (OutputEM). 
However, as a variety of examples produce outputs in numbers, we convert all numbers in \texttt{string} type to the \texttt{float} type with two decimal spaces to better match numbers.
Similarly, we remove the explanation string when printing outputs for better comparison.

\begin{table}[hbp]
  \centering
    \resizebox{0.45\textwidth}{!}{
    \begin{tabular}{lrl}
    \toprule
    Output Type & \multicolumn{1}{l} \%  & Examples  \\
    \midrule
    Single number & 55  & 0.841076906802073; 68 \\
    List/tuple/array & 34  & (256, 10); [``UserID", ``Gender"]  \\
    \multirow{2}[1]{*}{Dataframe} & \multirow{2}[1]{*}{11} & \quad Weight \, 26.25 \\
          &       & \quad Speed \,\,\,\, 36.70 \quad dtype: float64 \\
    \bottomrule
    \end{tabular}%
    }
  \caption{Types of ground truth outputs in ExeDS.}
  \label{tab:output-type}
\end{table}%

\section{Evaluating Code Generation on ExeDS}
Based on ExeDS, we evaluate the models' performance on data science code generation and compare both surface-form code and execution output.

\paragraph{DS Code Generation}
We investigate the task of target code cell generation in notebooks with context. 
Figure \ref{fig:example} presents an example of the task. 
For each target code cell, we prepare a source-target example, conditioned on prior multimodal context and natural language intent.  
The context includes: (1) the closest three cells prior to the target cell, regardless of code or markdown; (2) a code statement to define the columns names of data in the format of \texttt{df.columns=['a', 'b']}.

\paragraph{Baseline Models}

We test five code generation models: 
(1) PyMT5 \cite{Clement2020PyMT5} is an encode-decoder transformer \cite{Vaswani2017Transformers} pretrained on Python corpus.
(2) JuPyT5 \cite{Chandel2022JuPyT5} is an encoder-decoder transformer pretrained on Jupyter notebooks with the code-infilling objective.
(3) CodeGPT and CodeGPT-adapted \cite{lu2021codexglue} are two GPT-style models \cite{Solaiman2019GPTstyle} pretrained on CodeSearchNet Python functions \cite{Husain2019CodeSearchNet}, where the former is trained from scratch and the latter is trained from GPT-2 checkpoint. 
(4) GPT-neo \cite{gptneo} is a GPT-style model pretrained on The Pile \cite{Gao2021ThePILE}, a dataset with a variety of text sources including 8\% GitHub code. 
We evaluate three GPT-neo models with different parameters, including 125M, 1.3B, and 2.7B.
(5) Codex \cite{Chen2021CodeX} is the state-of-the-art model trained on 159G GitHub Python files from GPT-3 \cite{Brown2020GPT3}. 
We test its zero-shot performance due to the inaccessibility of model weights.

\paragraph{Finetuning} For training and validation, we filter a set of 123K source-target examples from JuiCe with data dependencies, where the target is any code cell and the source is the prior multimodal context as in ExeDS. 
We randomly select 4K examples for validation and leave the rest for finetuning. 
More details can be found in Appendix \ref{app-sec:hyper-param}.

\paragraph{Metrics} We report results with OutputEM, which is the proportion of examples with correct output, and surface-form metrics, i.e. BLEU, CodeBLEU \cite{Ren2020CodeBLEU}, and Exact Match (EM).

\begin{table}[tbp]
  \centering
    \resizebox{0.45\textwidth}{!}{
    \begin{tabular}{lccc|c}
    \toprule
          & \multicolumn{1}{c}{BLEU}  & \multicolumn{1}{c}{CodeBLEU} &  \multicolumn{1}{c|}{EM} &  \multicolumn{1}{c}{OutputEM} \\
    \midrule
    \multicolumn{3}{l}{\textit{\quad GPT-style framework}} & & \\ 
    GPT-neo-125M & 3.4    & 17.2  & 0.0    & 1.5   \\
    GPT-neo-1.3B & 9.2    & 26.2  & 0.0    & 10.7   \\
    GPT-neo-2.7B & 9.1    & 28.8  & 0.4    & 13.3   \\
    CodeGPT      & 26.4   & 28.6  & 1.5    & 12.7  \\
    CodeGPT-adapted & 25.1  & 26.8  & 3.3     & 13.1  \\
    Codex* & 3.9  & 23.5  & 0.0     & 27.7  \\
    \midrule
    \multicolumn{3}{l}{\textit{\quad encoder-decoder framework}} & & \\
    PyMT5 & 25.7  & 35.8  & 2.8    & 19.7 \\
    JuPyT5 & \textbf{35.3} & \textbf{41.1} & \textbf{6.2} & \textbf{31.6} \\
    \bottomrule
    \end{tabular}%
    }
  \caption{Evaluation results of surface form metrics and execution metric.
  * denotes a zero-shot setting.  }
  \label{tab:main-result}%
\end{table}%

\section{Evaluation Results}
In this section, we show and analyze evaluation results to show the advantages of our ExeDS dataset.

\subsection{Main Results}
Table \ref{tab:main-result} shows the results of different baseline models in surface form metrics and execution correctness. We have the following main observations.

(1) For all models, the surface form EM is close to zero while the OutputEM is in a normal range.
This suggests that surface form EM often fails to evaluate code correctness, while the execution metric is better which covers more correct cases and shows correctness beyond matching code strings. 

(2) Surprisingly, zero-shot Codex achieves compatible results with finetuned JuPyT5 in OutputEM, but it performs badly with surface-form metrics. 
This finding suggests the strength of Codex to generate correct code and understand the multimodal context.
In addition, the difference between surface-form scores and OutputEM again shows the superiority of measuring code with execution correctness.   

(3) Encoder-decoder models perform better than GPT-style models with all metrics, which indicates their strength in generating code. Also, JuPyT5 achieves the best performance with all metrics. One possible reason is that JuPyT5 is pretrained on a large corpus of notebooks, which learns the necessary knowledge from the notebook context.

\subsection{Error Analysis}\label{sec:error-analysis}
We give two error analyses of execution results to investigate examples with raised execution exceptions and erroneous outputs.
The code examples are produced by our top-performing model JuPyT5. 
Detailed examples can be found in Appendix \ref{app-sec:examples}.

\begin{table}[tbp]
  \centering
    \resizebox{0.475\textwidth}{!}{
    \begin{tabular}{lrr}
    \toprule
    Error Category & \multicolumn{1}{l}{\%} & \multicolumn{1}{l}{Exception Examples} \\
    \midrule
    Use undefined variable  & \multicolumn{1}{l}{45} & \multicolumn{1}{l}{NameError…} \\
    Use undefined API & \multicolumn{1}{l}{16} & \multicolumn{1}{l}{AttributeError…} \\
    Use wrong schema & \multicolumn{1}{l}{22} & \multicolumn{1}{l}{KeyError, ValueError, IndexError…} \\
    Wrong Syntax & \multicolumn{1}{r}{8} & \multicolumn{1}{l}{IndentationError, SyntaxError …} \\
    Other errors & \multicolumn{1}{r}{9} & \multicolumn{1}{l}{No message, ImportError, …} \\
    \bottomrule
    \end{tabular}%
    }
  \caption{Qualitative error analysis on examples that raise exceptions during execution. Some representative exception types for each error category are listed.      }
  \label{tab:error-analysis-exception}%
\end{table}%

\begin{table}[tbp]
  \centering
    \resizebox{0.34\textwidth}{!}{
    \begin{tabular}{lrl}
    \toprule
    Error Category & \multicolumn{1}{l}{\%} & Examples \\
    \midrule    
    Incorrect Code  & \multicolumn{1}{l}{56} & Figure \ref{app-fig:example11} \& \ref{app-fig:example12} \\
    No Output & \multicolumn{1}{r}{8} & Figure \ref{app-fig:example5}  \\ 
    Partially Correct & \multicolumn{1}{l}{12} & Figure \ref{app-fig:example6}  \\ 
    To Many Output & \multicolumn{1}{l}{24} & Figure \ref{app-fig:example9} \& \ref{app-fig:example10} \\ 
    \bottomrule
    \end{tabular}%
    }
  \caption{Analysis of 50 examples with wrong outputs. }
  \label{tab:error-analysis-output}%
\end{table}%

\begin{figure}[th]
     \centering
     \includegraphics[width=0.475\textwidth]{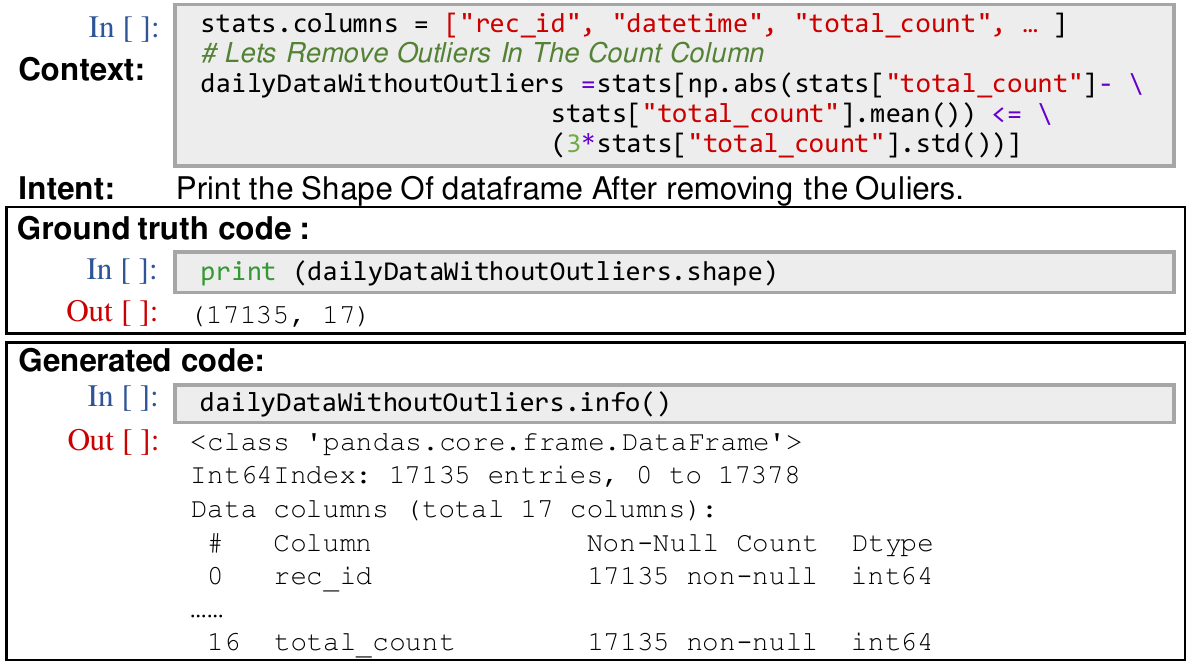}
     \caption{An incorrect example with high surface form metrics scores but low execution metrics scores. Surface form metrics are deficient to evaluate code correctness. }
     \label{fig:case-study-high-BLEU-low-Exe}
\end{figure}

\paragraph{Exception Types} Table \ref{tab:error-analysis-exception} shows five exception types from 154 examples. 
We find for 45\% cases, the model fails to capture data-flow and uses undefined variables in context.
For 16\% cases, the model misuses API methods and often leads to \texttt{AttributeError}
, possibly due to version differences and calling methods without import.  
22\% cases misuse the data schema of dataframes, which indicates the need to improve code generation models with such multimodal context, especially how to incorporate the data schema context. 
Only 8\% cases have syntax problems, suggesting the model's strong ability to generate syntax-correct code.  

\paragraph{Output Errors} Table \ref{tab:error-analysis-output} shows four types of output errors from 50 examples. 
We find 56\% cases have incorrect code.
The challenging NL description and context might be hard for models to understand and generate correct code.
8\% cases complete the correct functions but do not call \texttt{print()} to output.
12\% of cases are partially correct, where the output mismatch is caused due to some missing details, for example, the absence of some parameters. 
Finally, 24\% cases produce too many outputs.

\section{Case Study}

We give an example predicted by JuPyT5 with a high BLEU score but erroneous outputs in Figure \ref{fig:case-study-high-BLEU-low-Exe}, to show the advantages of execution evaluation for DS code generation.
The example is a typical DS task which intends to explore the shape of a dataframe.
But the model misunderstands the intents and generates code to display all dataframe information. 
Although we can find the expected shapes from the output, i.e., 17135 entries and 17 columns, the output is not exactly correct. 
However, as the code is short while the variable name is long, which leads to a high overlap between prediction and ground truth, the generated code obtains above average BLEU and CodeBLEU scores. 
This example reveals the deficiency of surface form metrics to evaluate code correctness.

\section{Conclusion}
In this paper, we propose an evaluation dataset to support execution correctness evaluation for data science code generation dubbed ExeDS, 
which consists of 534 typical data science problems from Jupyter Notebooks, each with code context, task description, target code, and desired execution output.
By performing experiments with five strong code generation models on ExeDS, we find models that achieve high surface-form scores do not necessarily produce execution correct code, and execution-based metrics could capture more detailed code generation errors. 
We expect our efforts to attract more attention to code execution correctness and generating executable code.

\section*{Limitations}
Firstly, only the test set examples have high quality of human annotation and verification. Thus the training set might be too noising to train a robust code generation model.
Secondly, the execution metric is insufficient to show other information like semantic relatedness, variable naming, and API usages, which are also important in evaluating a good code. 
Thirdly, our datasets and metrics focus on Python code in data science domain. It's unclear whether is applicable to general software code. 
Fourth, our execution-based automatic evaluation is more time-consuming to compute and evaluate than other surface-form metrics like EM, BLEU. 
At last, evaluating generated code is far different from evaluating natural languages. The final goal of code generation is to generate execution and functional correct code. Though with many limitation, our work could be a pilot study which provides insights and possible solutions on how to better evaluate code generation models.

\bibliography{anthology,custom}
\bibliographystyle{acl_natbib}

\newpage

\appendix

\section{Finetuning Details}\label{app-sec:hyper-param}

We finetune all the baseline models, except Codex, on our cleaned training set and select the best checkpoint with the perplexity score on dev set for testing. 
All models are trained on 16 Tesla V100 32GB GPUs. 
The hyper parameter are presented in Table \ref{app-tab:hyper-param}.

At inference time, we use beam search decoding with a beam size of 5. 

\begin{table*}[tbp]
  \centering
    \resizebox{0.98\textwidth}{!}{
    \begin{tabular}{l|c|c|c|c|c|c|c}
    \hline
    Hyperparameter & CodeGPT & CodeGPT-adapted & GPT-neo 125M & GPT-neo 1.3B & GPT-neo 2.7B & PyMT5 & JuPyT5 \\
    \hline
    \# vocab size & 50001 & 50260 & 50257 & 50257 & 50257 & 50337 & 50340 \\
    \# parameters & 124M  & 124M  & 125M  & 1.3B  & 2.7B  & 374M  & 374M \\
    \# hidden size & 768   & 768   & 768   & 2048  & 2560  & 1472  & 1472 \\
    \# layers & 12    & 12    & 12    & 16    & 20    & 12    & 12 \\
    \# heads & 12    & 12    & 12    & 24    & 32    & 12    & 12 \\
    dropout & 0.1   & 0.1   & 0.1   & 0.1   & 0.1   & 0.1   & 0.1 \\
    optimizer & AdamW & AdamW & AdamW & AdamW & AdamW & Adam  & Adam \\
    learning rate & 5e-05 & 5e-05 & 5e-05 & 5e-05 & 5e-05 & 1e-4 & 1e-4 \\
    batch size & 16    & 16    & 3     & 1     & 1     & 1     & 1 \\
    epochs & 30    & 30    & 10    & 10    & 10    & 10    & 10 \\
    max tokens & 512   & 512   & 2048  & 2048  & 1536  & 3600  & 3600 \\
    \hline
    \end{tabular}
    }
  \caption{Details of the hyperparameters used during fine-tuning for the code generation task in this paper.}
  \label{app-tab:hyper-param}
\end{table*}%

\section{More Examples}\label{app-sec:examples}

\begin{figure}[h]
     \centering
     \includegraphics[width=0.475\textwidth]{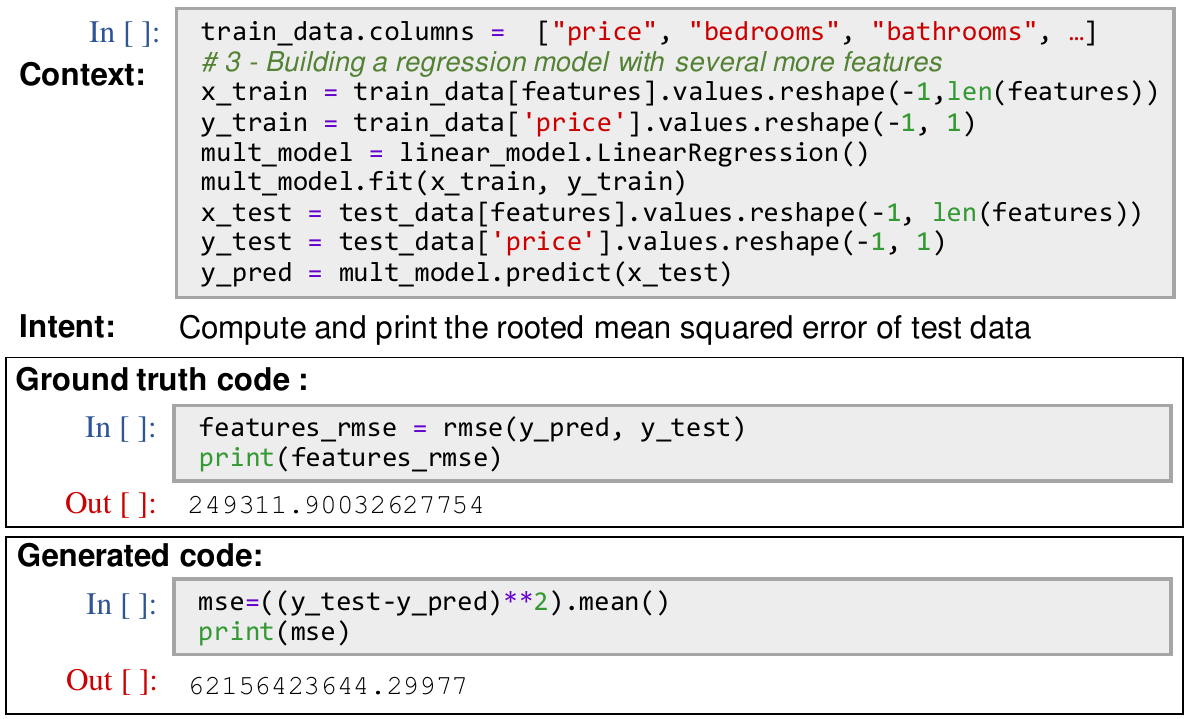}
     \caption{An example with incorrect code. The NL intent is too challenging and the generated code misses the key information to compute the \texttt(rooted) error. More powerful models to understand NL intent are required. }
     \label{app-fig:example11}
\end{figure}

\begin{figure}[h]
     \centering
     \includegraphics[width=0.475\textwidth]{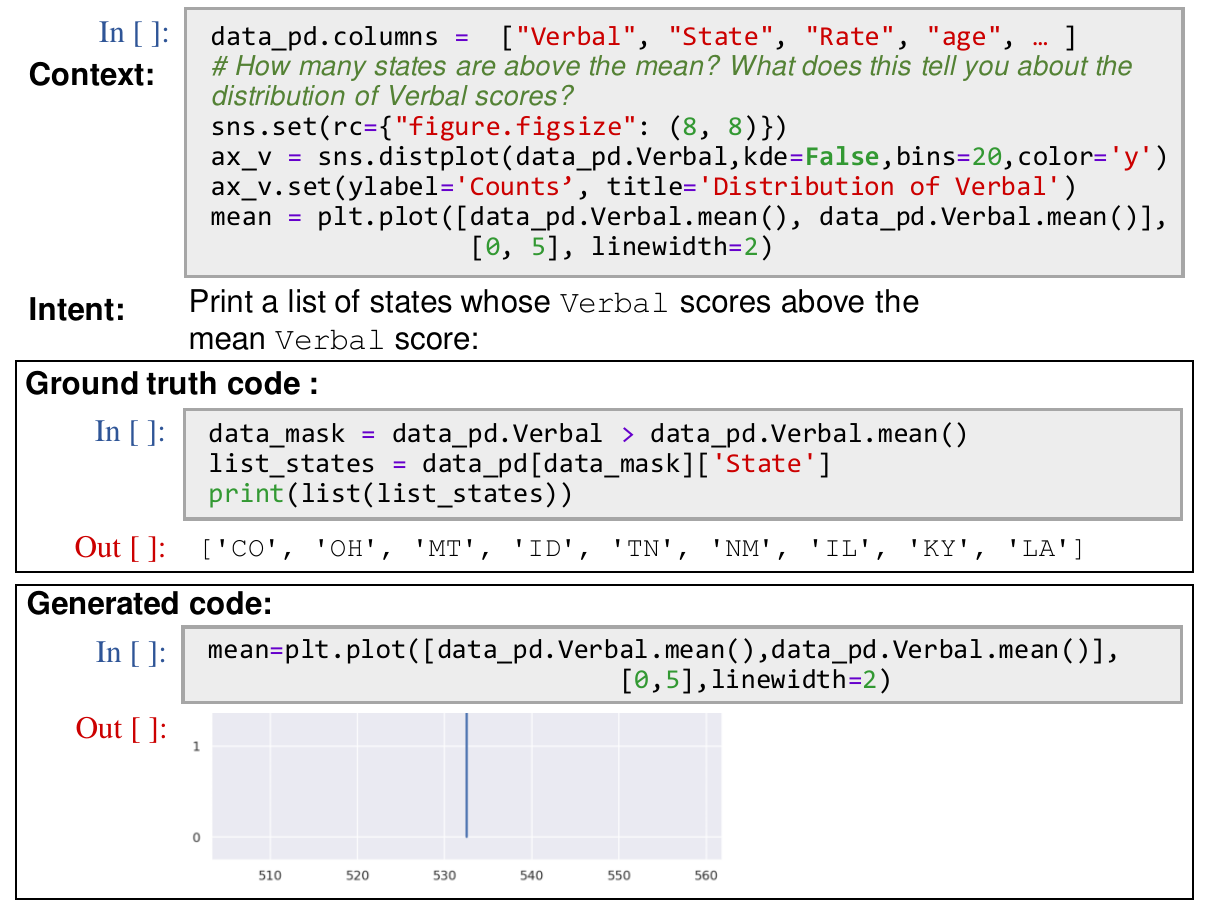}
     \caption{An example with incorrect code. The model fails to perform contextual reasoning over such multimodal context. }
     \label{app-fig:example12}
\end{figure}

\begin{figure}[h]
     \centering
     \includegraphics[width=0.475\textwidth]{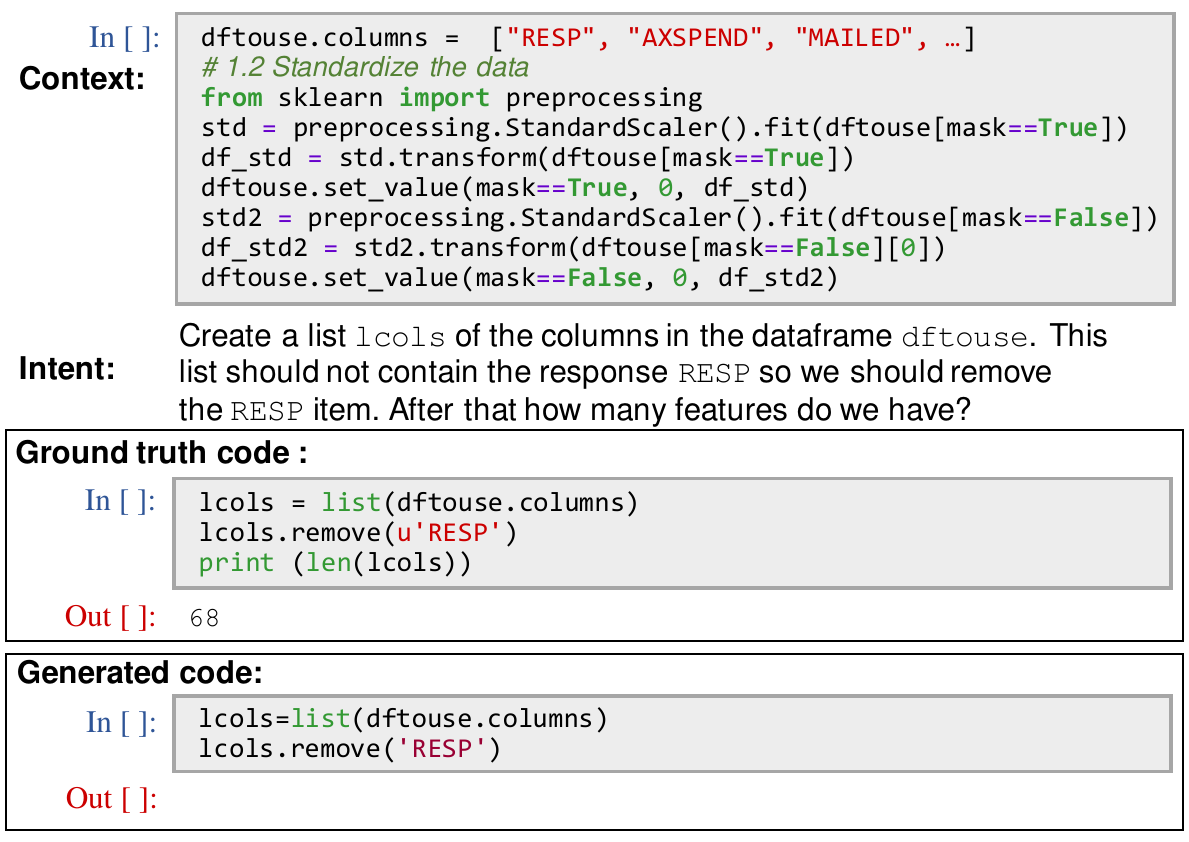}
     \caption{An example with a no output error.  the generated code satisfies the intent to create a list and remove the item. But it fails to produce the output, i.e., the length of the feature list. }
     \label{app-fig:example5}
\end{figure}

\begin{figure}[h]
     \centering
     \includegraphics[width=0.475\textwidth]{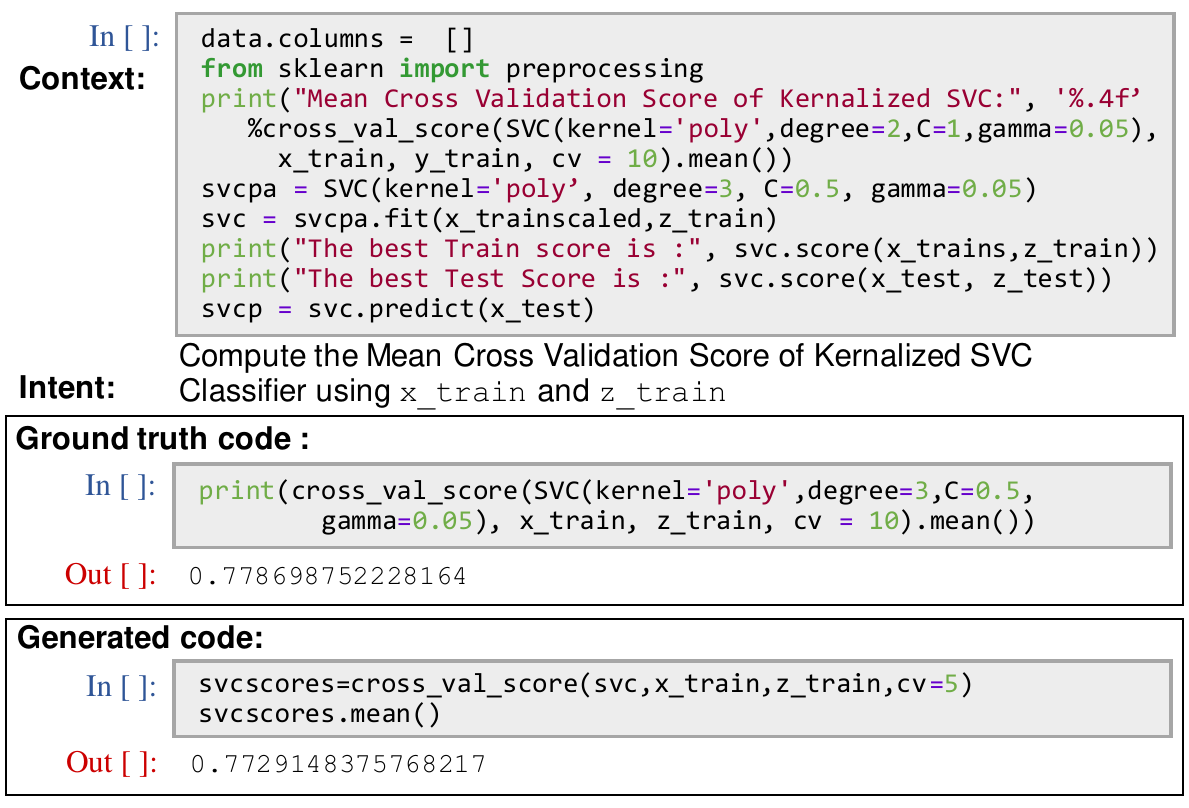}
     \caption{An example with a partially correct error. The code is actually correct but the parameter for \texttt{cv} is different, resulting in the difference between the ground truth and execution outputs.}
     \label{app-fig:example6}
\end{figure}

\begin{figure}[t]
     \centering
     \includegraphics[width=0.475\textwidth]{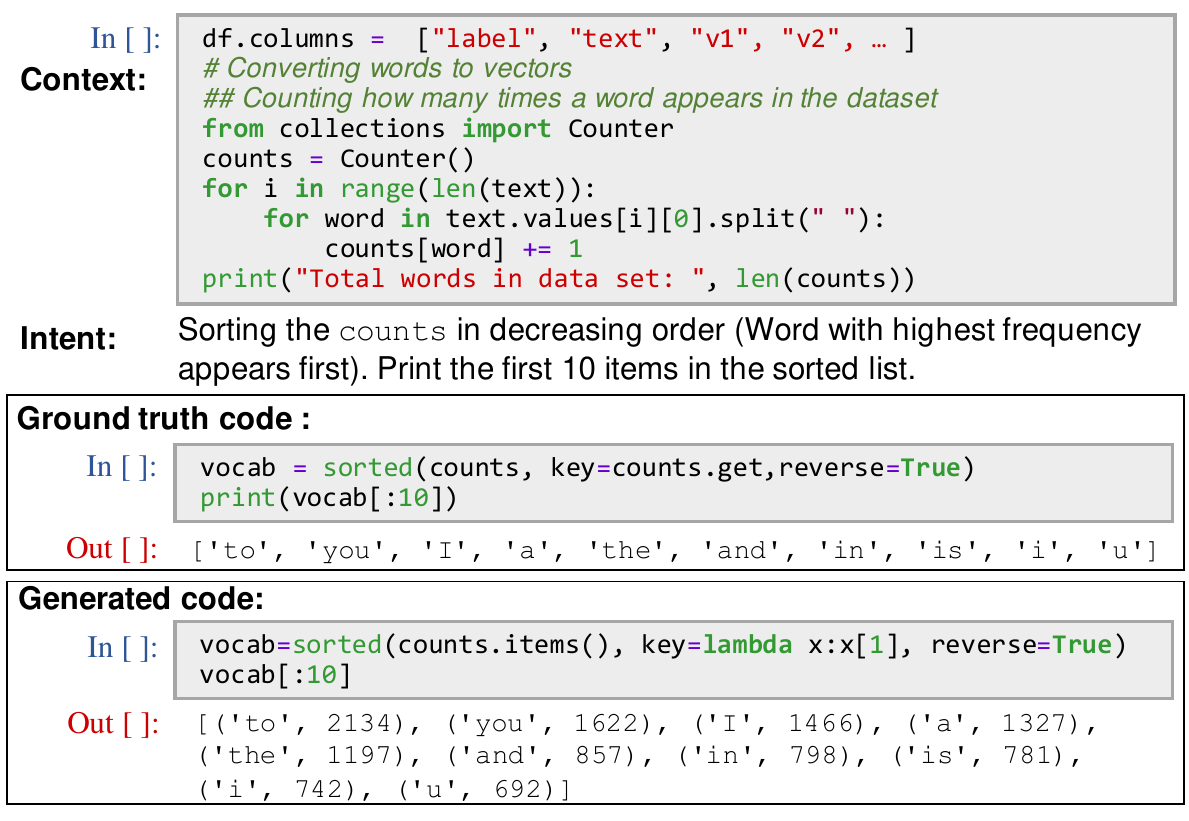}
     \caption{An example with too many output. The correct output actually exists in the execution output, but the excessive output causes the inexact match and decline in ExeF1. }
     \label{app-fig:example9}
\end{figure}

\begin{figure}[t]
     \centering
     \includegraphics[width=0.475\textwidth]{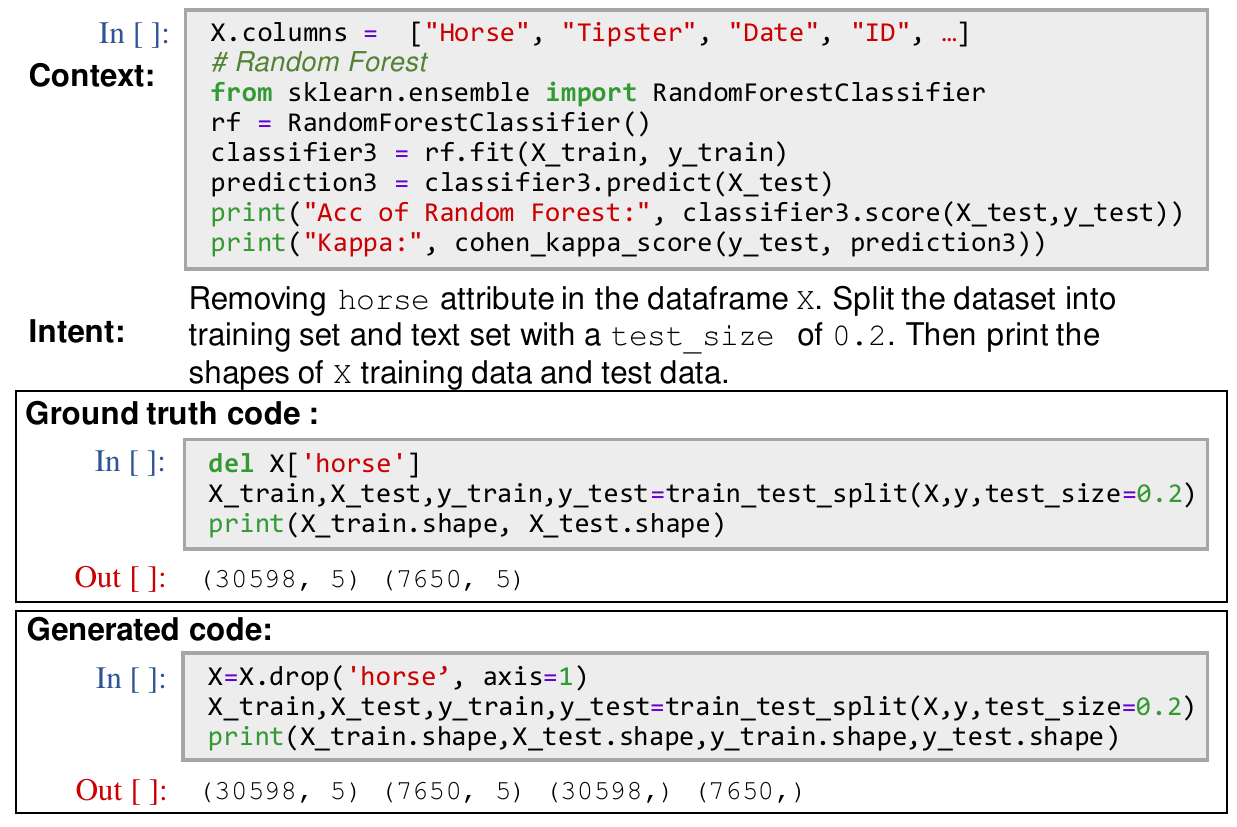}
     \caption{Another example with too many output.}
     \label{app-fig:example10}
\end{figure}

\begin{figure}[t]
     \centering
     \includegraphics[width=0.475\textwidth]{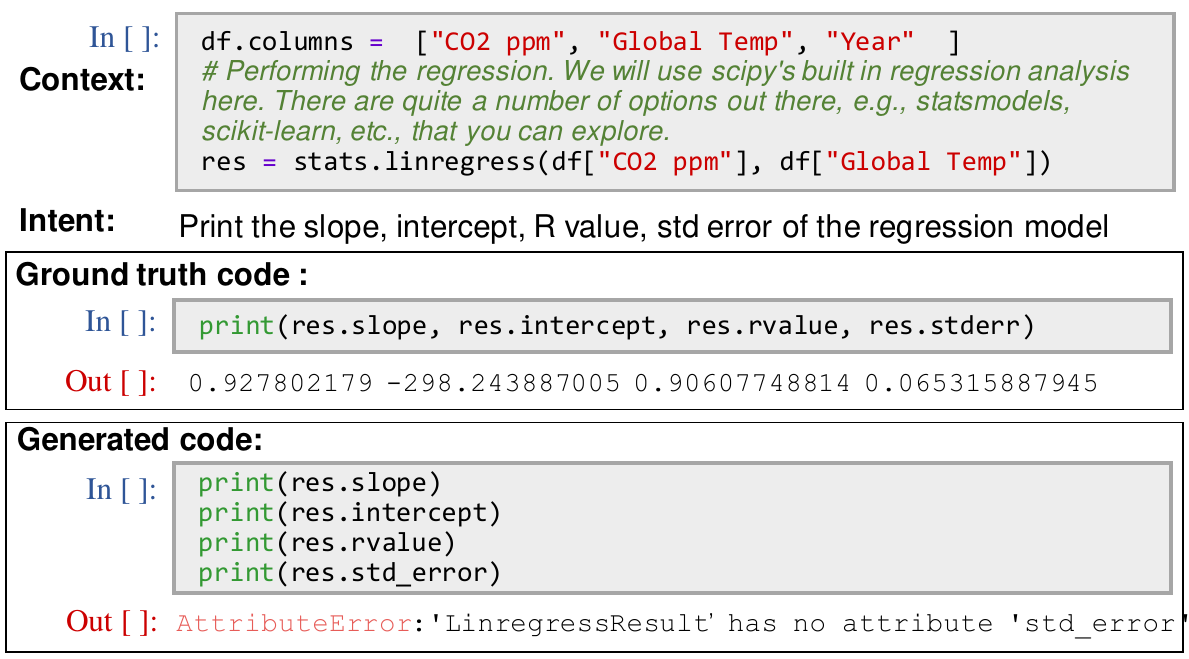}
     \caption{An example running with exceptions. The model misuses the attribute to call the standard errors.}
     \label{app-fig:example8}
\end{figure}

In this section, we present 6 examples to show the typical types of errors with erroneous outputs in Figure \ref{app-fig:example11} - Figure \ref{app-fig:example10}. 
We also give an example with a typical type of errors causing exceptions in Figure \ref{app-fig:example8}.

\end{document}